# Mechanisms of inductively coupled $BCl_3$-plasma interaction with the GaN surface

*A. A. Kobelev[1], N. A. Andrianov[2], Yu. V. Barsukov[1], and A. S. Smirnov[1]*
[1]Peter the Great Saint-Petersburg Polytechnic University 29 Politekhnicheskaya str., Saint-Petersburg, 195251, Russia
E-mail: kobelev_anton@mail.ru; barsukov.yuri@gmail.com; a._s._smirnov@mail.ru
[2]Svetlana-Rost, JSC 27 Engels av., Saint-Petersburg, 194156, Russia E-mail: n.andrianov@svrost.ru

*Consideration is given to inductively coupled $BCl_3$-plasma (ICP) treatment of the GaN surface, which is a promising technique to get the low resistance ohmic contacts in GaN-based transistors. In some cases, $BCl_3$ plasma treatment results in ohmic contact degradation because $BCl_x$ radicals tend to form a polymer thin film $B_xCl_y$ on the surface. In the present work, the mechanisms of $BCl_3$ plasma interaction with the GaN surface are considered. Threshold ion energies of reactive ion etching for polymer $B_xCl_y$ and semiconductor GaN, respectively, are estimated using numerical plasma modeling. It has been demonstrated that a plasma treatment regime without polymer deposition and reactive etching is possible when an ion energy is in the range ~32-60 eV.*

## Introduction

Transistors with high electron mobility (HEMT) based on wide-bandgap GaN compounds are promising radio-electronic components for use in high-power radars, communication, and telecommunication systems. The high breakdown voltage of GaN, combined with the high current density in the HEMT channel cross-section, allows for achieving higher power density in GaN-HEMT-based amplifiers compared to their GaAs-based counterparts [1]. The high thermal conductivity coefficient of 130 W/(m·K) (at 300 K) and the melting temperature of GaN, exceeding 2500 K, allow the operating temperature range of GaN-HEMT to be extended up to $T_{max} \approx 700$ °C, more than twice that of $T_{max} \approx 300$ °C for GaAs-HEMT [1].

The formation of ohmic contacts (OC) with low resistance is one of the key requirements in the production of GaN/AlGaN HEMT-based integrated circuits. Processing the GaN surface with zero etch rate in the plasma of a radio-frequency inductively coupled plasma (RF ICP) gas discharge in $BCl_3$ gas before the metal deposition stage is a promising method for obtaining OCs with low resistance and high uniformity across the substrate area [2]. The main advantage of this method is the elimination of the need for highly accurate control over the etching depth and width profile across the entire substrate area. The absence of etching is ensured by reducing the average energy of the ions bombarding the material to a value close to the floating potential of the RF ICP plasma discharge.

The reduction of OC resistance after $BCl_3$ plasma treatment is ensured by two main factors [2, 3]. The first factor is the reduction of nitrogen content relative to gallium on the GaN surface under the influence of ion bombardment from the plasma, which leads to a decrease in the potential barrier at the "metal–semiconductor" interface. Secondly, the use of RF ICP discharge in $BCl_3$ gas allows for the removal of oxygen from the GaN surface, thereby cleaning the contact area between the semiconductor and the deposited metal. The cleaning property of $BCl_3$ discharges is explained by the ability of $BCl_x$ radicals, x = 1 – 2, as the main dissociation products in the plasma, to form chemical bonds with oxygen and create stable volatile compounds of boron oxychloride such as $(BOCl)_3$, $B_3O_3Cl_2$, $B_2OCl_3$, and others, which are removed from the surface by ion bombardment [4]. Due to this property, gas discharges in a $BCl_3$ mixture have found widespread application both for cleaning surfaces from oxide films [5–7] and in reactive ion etching technologies for materials based on compounds with oxygen ($PbZr_xTi_{1-x}O_3$, $SiO_2$, $ZnO_2$, $HfO_2$, $ZrO_2$, $Zr_{1-x}Al_xO_y$) [8–12].

However, $BCl_x$ radicals tend to deposit and form a $B_xCl_y$-type polymer on the surface of solids, which blocks the access of ions and radicals to the surface [11, 13, 14]. Consequently, the etching process of the main material completely stops and is replaced by the deposition of the polymer on its surface. In the case of treating the GaN surface in an RF ICP $BCl_3$ plasma discharge, a $B_xCl_y$ polymer film may form instead of removing the Ga–O surface oxide layer, resulting in an additional contaminating layer at the "metal–semiconductor" contact interface [15]. This leads to a significant increase in OC resistance.

This work is devoted to the numerical modeling of $BCl_3$ plasma RF ICP discharge and its interaction with the GaN surface. The transition from the deposition mode of the $B_xCl_y$ polymer film to the etching mode of GaN, depending on the amplitude of the RF ICP voltage applied to the substrate with the sample, in the range from 0 to 150 V, is demonstrated. A comparison of the numerical calculation data with experimental results [3, 15] was carried out, and a model for the reactive ion etching of GaN in $BCl_3$ plasma was developed.

## Numerical Modeling

Figure 1 shows a schematic representation of the Corial 210D experimental setup, which was previously used in studies [3, 15] to investigate the effect of RF ICP plasma discharge of $BCl_3$ treatment on the resistance of OC formed on the surface of GaN/AlGaN epitaxial heterostructures. The sample was placed on the lower RF ICP electrode, which was connected to an RF generator with a frequency of $f_1$ = 13.56 MHz. In the study [3], during plasma treatment, the power of this RF generator was adjusted to fix the self-bias voltage $U_{dc}$ on the lower electrode in the range from 0 to -60 V. The power of the RF ICP discharge at 200 W was controlled using a solenoid coil with an RF current at a frequency of $f_2$ = 2 MHz. Four turns of this coil encircle the main volume of the discharge chamber, which has dielectric sidewalls. The neutral gas pressure in the chamber was 10 mTorr in all measurements. The $BCl_3$ gas was introduced through holes in the upper grounded electrode. The diameter of the discharge chamber is 26 cm, and the distance between the upper and lower electrodes is 17.3 cm.

The experimental parameters were used as input data for the numerical calculation of the spatial distribution of $BCl_3$ plasma parameters. Within the framework of the hydrodynamic approximation, numerical solutions of the continuity equations, force balance, and energy balance, along with Maxwell's equations, were performed using the commercial code CFD-ACE+ [16]. The problem was solved in a two-dimensional setup with axial symmetry. The axis of axial symmetry in the model corresponds to the line passing through the centers of the lower RF ICP electrode and the upper grounded electrode (Figure 1).

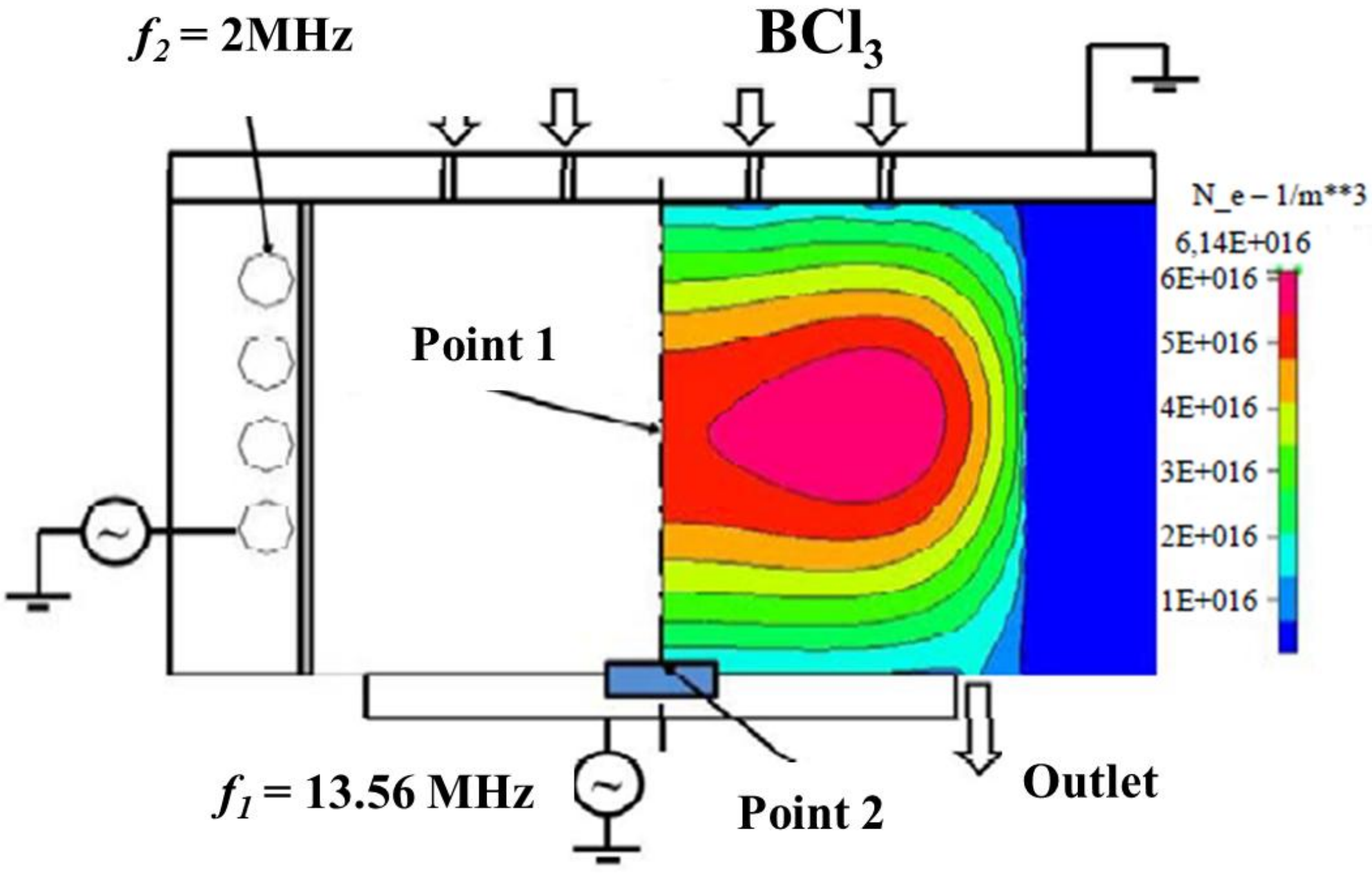

**Figure 1**. Schematic illustration of the discharge chamber geometry and the calculated distribution of the electron concentration averaged over the RF period $T_{RF} = f_2^{-1}$ in the plasma of the RF discharge at an RF-voltage amplitude of $U_{rf} = 0\ V$.

The boundary condition for the electric potential on the RF electrode is given by: $\varphi = U_{DC} + U_{rf} \times \cos{(2\pi f \times t)}$, where $U_{rf}$ is the amplitude of the RF voltage at a frequency of 13.56 MHz, $U_{dc}$ is the self-bias voltage, which is automatically calculated during the simulation. The value of $U_{rf}$ varied in the range from 0 to 150 V. The electric potential on the upper grounded electrode was fixed at $\varphi = 0V$.

To describe the $BCl_3$ plasma, literature data on ionization cross-sections, excitation of vibrational and electronic levels by direct electron impact for $BCl_3$ molecules, and dissociation products $BCl_{x=0–2}$, and Cl were used [17, 18]. Additionally, processes of negative ion $Cl^-$ formation, volumetric recombination of positive and negative ions, and non-resonant charge exchange of ions were included [17, 18]. The complete set of processes considered in the description of BCl3 plasma is presented in Table 1.

**Table 1**. List of reactions describing $BCl_3$ RF ICP discharge.

| No. | Reaction | Description |
|---|---|---|
| 1 | $BCl_3 + e \rightarrow BCl_3 + e$ | Elastic scattering of electrons |
| 2 | $BCl_3 + e \rightarrow BCl_3(v) + e$ | Vibrational excitation |
| 3 | $BCl_3 + e \rightarrow BCl_2 + Cl + e$ | Dissociation by electron impact |
| 4 | $BCl_3 + e \rightarrow BCl + Cl_2 + e$ | Dissociation by electron impact |
| 5 | $BCl_3 + e \rightarrow BCl_2 + Cl^-$ | Dissociative electron attachment |
| 6 | $BCl_3 + e \rightarrow BCl_2^+ + Cl + 2e$ | Dissociative ionization |
| 7 | $BCl_3 + e \rightarrow BCl^+ + Cl_2 + 2e$ | Dissociative ionization |
| 8 | $BCl_3 + e \rightarrow BCl_2 + Cl^+ + 2e$ | Dissociative ionization |
| 9 | $Cl_2 + e \rightarrow Cl_2^+ + 2e$ | Ionization |
| 10 | $Cl_2 + e \rightarrow Cl^- + Cl$ | Dissociative electron attachment |
| 11 | $Cl_2 + e \rightarrow 2Cl + e$ | Dissociation |
| 12 | $Cl^- + e \rightarrow Cl + 2e$ | Electron detachment |
| 13 | $Cl + e \rightarrow Cl^+ + 2e$ | Ionization |
| 14 | $Cl + e \rightarrow Cl^* + e$ | Electronic excitation |
| 15 | $Cl^* + e \rightarrow Cl^+ + 2e$ | Multi-step ionization |
| 16 | $BCl_2 + e \rightarrow BCl_2(v) + e$ | Vibrational excitation |
| 17 | $BCl + e \rightarrow BCl(v) + e$ | Vibrational excitation |
| 18 | $BCl_2 + e \rightarrow BCl + Cl + e$ | Dissociation by electron impact |
| 19 | $BCl + e \rightarrow B + Cl + e$ | Dissociation by electron impact |
| 20 | $BCl_2 + e \rightarrow BCl + Cl^- + e$ | Dissociative electron attachment |
| 21 | $BCl + e \rightarrow B + Cl^- + e$ | Dissociative electron attachment |
| 22 | $B + e \rightarrow B + e$ | Electronic excitation |
| 23 | $Cl_2^+ + Cl– \rightarrow Cl_2 + Cl$ | Ion-ion recombination |
| 24 | $Cl^+ + Cl^- \rightarrow Cl + Cl$ | Ion-ion recombination |
| 25 | $BCl_3^+ + Cl^- \rightarrow BCl_3 + Cl$ | Ion-ion recombination |
| 26 | $BCl_2^+ + Cl– \rightarrow BCl_2 + Cl$ | Ion-ion recombination |
| 27 | $BCl^+ + Cl^- \rightarrow BCl + Cl$ | Ion-ion recombination |
| 28 | $Cl_2 + Cl^+ \rightarrow Cl_2^+ + Cl$ | Non-resonant charge transfer |
| 29 | $BCl_3 + Cl^+ \rightarrow BCl_2^+ + 2Cl$ | Non-resonant charge transfer |
| 30 | $BCl_3 + BCl^+ \rightarrow BCl_2^+ + BCl + Cl$ | Non-resonant charge transfer |
| 31 | $B + Cl^- \rightarrow BCl + e$ | Electron detachment |
| 32 | $BCl_2 + Cl^- \rightarrow BCl_3 + e$ | Electron detachment |
| 33 | $B + Cl_2 \rightarrow BCl + Cl$ | Radical recombination |
| 34 | $BCl + Cl_2 \rightarrow BCl_2 + Cl$ | Radical recombination |
| 35 | $BCl_2 + Cl_2 \rightarrow BCl_3 + Cl$ | Radical recombination |
| 36 | $Cl^* + Cl_2 \rightarrow 3Cl$ | Dissociation |

The electron mobility was calculated based on the transport cross-section of collisions between electrons and neutral $BCl_3$ particles [17]. For the calculation of the transport cross-section and ion mobility, data on the polarizability of $BCl_{x=0–3}$ and Cl atoms and molecules were used [18].

The boundary condition for charged particles was set as zero flux to the walls. The recombination of $BCl_x$ and Cl radicals was specified in the form of surface chemical reactions, presented in Table 2 [18]. The states of elements in the gas phase and on the surface of a solid are denoted by the indices (g) and (s), respectively. "Wall" denotes the quartz or aluminum wall.

**Table 1**. List of reactions describing $BCl_3$ RF ICP discharge.

| No. | Reaction | Probability |
|---|---|---|
| 1 | $BCl_{x=0–3}$(g) + wall → $BCl_{x=0–3}$(g) + wall | 1.00 |
| 2 | $Cl_{x=1,2}$(g) + wall → $Cl_{x=1,2}$(g) + wall | 1.00 |
| 3 | Cl*(g) + wall → Cl(g) + wall | 1.00 |
| 4 | Cl(g) + wall → Cl(s) | 1.00 |
| 5 | Cl(g) + Cl(s) → $Cl_2$(g) + wall | 0.05 |
| 6 | $BCl_2$(g) + wall → $BCl_2$(s) | 0.10 |
| 7 | BCl(g) + wall → BCl(s) | 0.20 |
| 8 | B(g) + wall → B(s) | 1.00 |
| 9 | Cl(g) + B(s) → BCl(s) | 0.10 |
| 10 | Cl(g) + BCl(s) → $BCl_2$(s) | 0.05 |
| 11 | Cl(g) + $BCl_2$(s) → $BCl_3$(g) + wall | 0.0005 |
| 12 | B(g) + $BCl_2$(s) → BCl(g) + BCl(s) | 0.05 |
| 13 | BCl(g) + $BCl_2$(s) → $BCl_2$(g) + BCl(s) | 0.02 |

## Results of the modeling at $U_{rf}$=0

The right part of Figure 1 shows the distribution of the period-averaged electron concentration at zero amplitude $U_{rf}$=0 on the RF electrode and self-bias voltage $U_{DC}$=0. The maximum plasma concentration $n_{pl}$=6.14×10$^{16}$ m$^{-3}$ is located near the side dielectric walls close to the RF current coils. For simplicity, the text below provides the concentration values of various components at point 1 in Figure 1, and the ion and radical fluxes to the substrate at point 2 in Figure 1, which correspond to the central area of the chamber near the axis of symmetry. The exclusion of the plasma parameter distribution inhomogeneity from consideration is justified by the fact that in the experiments [3, 15], samples of about 5 cm in size were used, which is much smaller than the 26 cm diameter of the discharge chamber.

The main positive ions are $BCl_2^+$ and $BCl_3^+$ with concentrations of $n(BCl_2^+)$ = 4.96×10$^{16}$ $m^{-3}$ and $n(BCl_3^+)$ = 1.93×10$^{16}$ $m^{-3}$. The concentrations of $BCl^+$ and $Cl^+$ ions are ~ (2−3)× 10$^{14}$ $m^{-3}$. The electron temperature $Te$ is 2.42 eV, calculated at point 1 in Figure 1. The electric potential drop between the plasma and the lower electrode is 15.4 V. his value is close to the floating potential $U_{fl}$≈12.7 $V$, which was estimated using the formula from [19]:

$$U_{fl} \cong \frac{T_e}{2e} \ln\left(\frac{M_i}{2.3m_e}\right), \quad (1)$$

where $M_i$ and $m_e$ are the mass of the main ion $BCl_2^+$ and the mass of the electron, respectively. Thus, at zero amplitude of the voltage on the RF electrode, ions from the ICP discharge plasma bombard the substrate with an average energy of no more than 15.4 eV.

The concentration of negative ions n($Cl^-$) = $1.56\times10^{16}\ m^{-3}$ is comparable to the concentration of electrons $n_e$=$5.4\times10^{16}\ m^{-3}$, which qualitatively agrees with experimental data [20]. Since negative ions with energies on the order of the neutral gas temperature cannot overcome the potential barrier and recombine on the chamber walls, the primary loss mechanism for $Cl^-$ ions is volumetric recombination with $BCl_2^+$ and $BCl_3^+$ in the positive column. The main radicals in $BCl_3$ plasma are Cl, BCl, and $BCl_2$ molecules with concentrations n(Cl)=$1.29\times10^{20}\ m^{-3}$, n(BCl) = $0.34\times10^{20}\ m^{-3}$, and n($BCl_2$) = $1.29\times10^{20}\ m^{-3}$, respectively.

## Results of the modeling at $U_{rf}\neq0$

Figure 2 shows the dependencies of the potential drop averaged over the RF period $T_{rf}$=$1/f_2$ in the sheath near the RF electrode, the plasma potential relative to the grounded electrode, and the total ion current as a function of $U_{rf}$ = 0 – 150 V.

The increase in the amplitude of the RF voltage $U_{rf}$ leads to a rise in the plasma potential from the floating potential value $<U_{pl}>=U_{fl}$=15.4 $V$ (Figure 2, black squares) to $<U_{pl}>$=46 $V$, and an increase in the self-bias voltage $U_{DC}$ from 0 to -88.8 V (Figure 2, red circles). The obtained curves qualitatively match the experimental data for argon [21] and $CF_4$ [22] under similar ICP conditions.

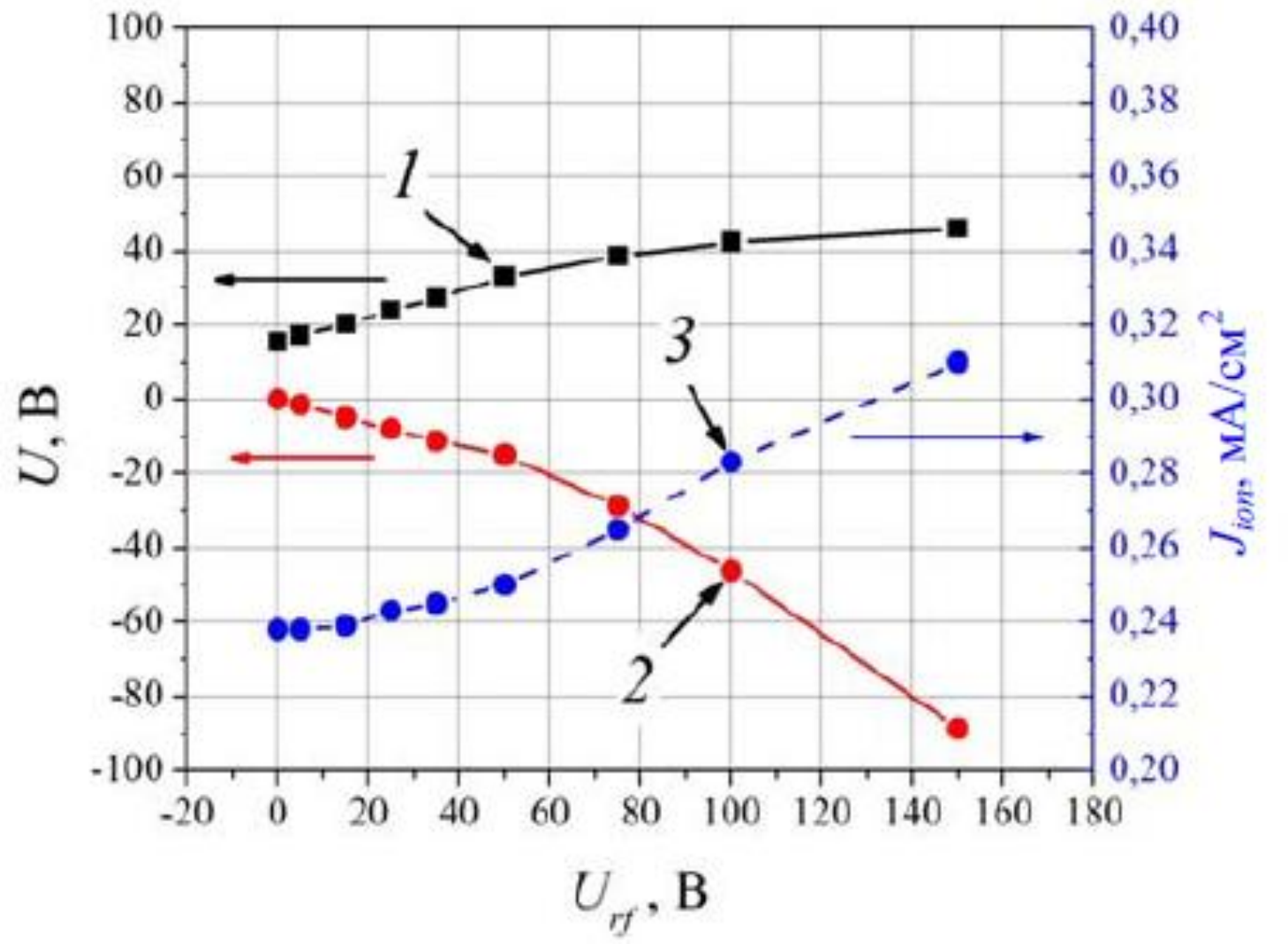

**Figure 3**. Calculation results for various values of the RF-voltage amplitude $U_{rf}$:1 – plasma potential averaged over the RF period relative to the grounded electrode; 2 – RF-period-averaged

sheath potential drop at the RF electrode $\langle U_{sh}\rangle$; 3 – total ion current density at the center of the lower electrode.

If we assume that the ions move through the sheath without collisions, the average energy of ions bombarding the GaN sample is calculated as $e<U_{sh}>$. This approximation is justified for RF discharges at pressures ≲ 10 mTorr [19, 22, 23]. Since ions enter the sheath at various phases of the electric field oscillations, the broadening of the ion energy spectrum is calculated according to [23, 24]:

$$\Delta E = \frac{e}{\pi f_1}\left(\frac{2e}{M_i}\right)^{\frac{1}{4}}\left(\frac{J_{ion}}{\varepsilon_0}\right)^{\frac{1}{2}}\left(U_{sh}^{p-p}\right)^{\frac{3}{4}}, \quad (2)$$

where $U_{sh}^{p-p}$ is the change in voltage in the sheath from minimum to maximum, and $J_{ion}$ is he total ion current density to the electrode considering all types of ions. The contribution of the main ion $BCl_2^+$ accounts for 70% of the total current.

The value of $Jion$ on the RF electrode changes by a factor of 1.29 as $U_{rf}$ increases from 0 to 150 V (Figure 2, blue circles). Thus, the amplitude of the RF voltage has a minimal impact on the ion current magnitude, which qualitatively agrees with numerical modeling data [25] and experimental measurements [26]. Similarly, the fluxes of neutral particles $BCl_x$ and Cl are minimally affected by the RF voltage amplitude [25]. For comparison, it is noted that increasing the ICP power from 200 to 400 W with $U_{rf}$=0 leads to a 2.9-fold increase in ion current and a 1.8 to 2.4-fold increase in the flux of $BCl_x$=1,2 and Cl radicals.

## Interaction of $BCl_3$ plasma and GaN surface

The GaN surface is typically covered with a Ga–O native oxide layer [2, 3]. This creates conditions for the deposition of a $B_xCl_y$ polymer film since $BCl_x$ radicals from the $BCl_3$ plasma can form a strong compound $(BOCl)_3$ with surface oxygen. If the ion energy $\varepsilon_{ion}$exceeds a certain threshold $\varepsilon_{trh}^{BxCly}$, ion bombardment results in the removal of surface oxygen in compounds such as $(BOCl)_3$, $B_3O_3Cl_2$, $B_2OCl_3$, $B_2OCl_4$, $B_2OCl_5$, and $B_2Cl_4$ [4]. he estimated threshold $\varepsilon_{trh}^{BxCly}$ ranges from 18 to 32 eV, according to experimental studies of the etching process in $BCl_3$ plasma for ZrAlO [8], $SiO_2$ [9], $HfO_2$ [9], and $ZrO_2$ [9]. Table 3 presents the results of experimental measurements of the etching rate (ER) and the composition of the GaN surface using X-ray photoelectron spectroscopy at different self-bias voltages $U_{DC}$= 0, −20, −40, $and$ −60 $V$ [3, 15].

**Table 3.** Experimental results: surface composition and etching rates.

| $meas.$, V | $ER^{meas.}$, nm/min | B, % | Cl, % | Ga, % | N, % | O, % | C, % |
|---|---|---|---|---|---|---|---|
| GaN before treatment | | - | - | 31.3 | 39.24 | 14.26 | 14.7 |
| 0 | $B_xCl_y$ polymer growth | 26.67 | 34.97 | 3.11 | 1.74 | 8.7 | 20.6 |
| -20 | ~0 | 4.25 | 2.64 | 26.91 | 42.55 | 13.33 | 10.33 |
| -40 | 1.8 | 1.89 | 1.45 | 31.18 | 34.49 | 16.2 | 10.64 |
| -60 | 3.0 | - | - | - | - | - | - |

After treatment in $BCl_3$ plasma at $U_{rf} = 0V$ and $U_{DC} = 0V$, a high ~ 30% content of B and Cl elements is observed, indicating the inevitable deposition of a $B_xCl_y$ polymer on the GaN surface [15]. Changing $U_{DC}$ to -20 V leads to a significant reduction in the percentage of B and Cl to ~2-4% and a noticeable increase in the surface content of Ga and N to ~27% and 43% respectively. During this process, the GaN etching rate remains zero [3]. Thus, the threshold energy $\varepsilon_{trh}^{BxCly}$, at which the removal of compounds like $(BOCl)_3$, $B_3O_3Cl_2$, $B_2OCl_3$, and others, as well as the $B_xCl_y$ polymer film, occurs, is lower than the threshold energy $\varepsilon_{trh}^{GaN}$ for etching GaN. The values of $\varepsilon_{trh}^{BxCly}$ and $\varepsilon_{trh}^{GaN}$ can be estimated by comparing the numerical calculation results with the experimental measurements from Table 3.

In Table 4, the results of the calculated period-averaged potential drop in the sheath near the lower electrode ($<U_{sh}>$), the average ion energy ($\varepsilon_{ion} = e<U_{sh}>$), and the energy spread ($\Delta E$), corresponding to the measured values of $U_{DC}$ from the experiment [3], are presented. The values of $<U_{sh}> = <U_{pl}> - U_{DC}$ were obtained from Figure 2, and the energy spread $\Delta E$ was determined according to formula (2), considering that $BCl_2^+$ ions are the main positive ions.

Since the polymer film is absent at $U_{DC} = -20V$ and $<\varepsilon_{ion}>=e<U_{sh}>=55\ eV$, the value of $\varepsilon_{trh}^{BxCly}$ lies in the range of 15.4 to 48.4 eV, considering the energy spread $\Delta E$=13.2 $eV$. This aligns with the results of $\varepsilon_{trh}^{BxCly} = 18{-}32\ eV$ obtained for ZrAlO [8], $SiO_2$ [9], $HfO_2$ [9], and $ZrO_2$ [9]. The etching of GaN in $BCl_3$ plasma starts at $U_{DC} > -20V$ [3]. Therefore, the value of $\varepsilon_{trh}^{GaN}$ ranges from 48.4 to 71 eV, taking into account the ion energy spread $\Delta E$.

**Table 4.** Comparison of Numerical Calculation Results with Experimental Measurements

| $U_{DC}^{meas}$,$V$ | $ER^{meas.}$,$nm/min$ | $U_{sh}^{simul}$,$V$ | $<\varepsilon_{ion}>$,$eV$ | $\Delta E$, $eV$ |
|---|---|---|---|---|
| 0 | $B_xCl_y$ growth | 15.4 | 15.4 | - |
| -20 | ~0 | 55 | 55 | 13.2 |
| -40 | 1.8 | 81 | 81 | 20 |
| -60 | 3.0 | 103.5 | 103.5 | 25.4 |

## Model of the reactive ion etching

Table 5 lists the main processes occurring on the GaN surface under the influence of ion and radical fluxes from the $BCl_3$ plasma. The indices (g) and (s) indicate the gas-phase and surface states, respectively. The following processes are considered in this model: chlorination of the surface (reaction $R_1$), polymer growth (reactions $R_2$ and $R_3$), etching of GaN (reaction $R_4$), and removal of the polymer along with surface oxygen (reaction $R_5$). The GaN surface is denoted as $\theta_0$, and the chlorinated GaN surface is denoted as $\theta_1$. Since the exact stoichiometric composition of the $B_xCl_y$ polymer is unknown, it is simplified in the model as a compound $BOCl_z$, and the polymer-covered surface is denoted as $\theta_2$. The total quantity $\theta_0+\theta_1+\theta_2$ is normalized to one.

**Table 5.** Reactions on GaN surface

| $R_i$ | Reaction | $\gamma i$ |
|---|---|---|
| Adsorption of Cl on GaN | | |
| $R_1$ | $Cl(g) + \theta_0(s) \rightarrow \theta_1(s)$ | 1 |
| Polymer growth | | |
| $R_2$ | $BCl_x(g) + \theta_0(s) \rightarrow \theta_2(s)$ | 0.0002 |
| $R_3$ | $BCl_x(g) + \theta_1(s) \rightarrow \theta_2(s)$ | 0.0002 |
| Reactive ion etching of GaN | | |
| $R_4$ | $BCl_x^+(g) + \theta_1(s) \rightarrow GaCl(g) + 0.5N_2(g) + BCl_x(g) + \theta_0(s)$ | $\gamma_{04}(\frac{\sqrt{\varepsilon_{ion}}}{\sqrt{\varepsilon_{trh}^{GaN}}} - 1)$ |
| Reactive ion etching of the polymer | | |
| $R_5$ | $BCl_x^+(g) + \theta_2(s) \rightarrow B_2OCl_{z+x}(g) + \theta_1(s)$ | $\gamma_{05}(\frac{\sqrt{\varepsilon_{ion}}}{\sqrt{\varepsilon_{trh}^{BxCly}}} - 1)$ |

The probability $\gamma_1$ of atomic chlorine deposition on GaN is assumed to be one [18]. Processes $R_2$ and $R_3$ occur with equal probabilities $\gamma_2=\gamma_3$, which is a fitting parameter in the model. To describe the reactive ion etching in reactions $R_4$ and $R_5$, the sputtering yield was used, which near the threshold is proportional to the square root of the average ion energy $\langle \varepsilon_{ion} \rangle$ [27, 28]:

$$\gamma_{4,5} = \gamma_{04,05}\left(\frac{\sqrt{<\varepsilon_{ion}>}}{\sqrt{<\varepsilon_{trh}>}} - 1\right), \quad (3)$$

where $\gamma_0$ is a fitting parameter based on experimental data on the GaN etching rate [3]. The threshold values were chosen as $\varepsilon_{trh}^{BxCly} \approx 31.9\ eV$ and $\varepsilon_{trh}^{GaN} \approx 59.7\ eV$. These values correspond to the midpoints of the previously mentioned ranges: $\varepsilon_{trh}^{BxCly}$ = 15.4–48.4 $eV$ and $\varepsilon_{trh}^{GaN}$ = 48.4–71 $eV$. Assuming that the surface composition does not depend on time, the following system of equations was compiled in accordance with the model presented in Table 5:

$$\theta_0 = 1 - \theta_1 - \theta_2, \quad (4)$$

$$\frac{1}{p_s}\frac{d\theta_1}{dt} = \gamma_1 J_{Cl}\theta_0 - \gamma_3 J_{BClx}\theta_1 - \gamma_4 J_{BClx+}\theta_1 + \gamma_5 J_{BClx+}\theta_2 = 0, \quad (5)$$

$$\frac{1}{p_s}\frac{d\theta_2}{dt} = \gamma_2 J_{BClx}\theta_2 + \gamma_3 J_{BClx}\theta_1 - \gamma_5 J_{BClx+}\theta_2 = 0, \quad (6)$$

where $p_s$ is the surface site density; $J_{Cl}$, $J_{BClx}$, and $J_{BClx+}$ are the fluxes of atomic chlorine, total flux of $BCl_x$ radicals, and total flux of $BCl_x^+$ ions obtained from the numerical calculation, respectively. The solution to the system of equations is given by the following expressions:

$$\theta_0 = \frac{\gamma_4\gamma_5 J_{BClx+}^2}{\gamma_2\gamma_3 J_{BClx}^2 + J_{BClx}(\gamma_2 J_{BClx+}(\gamma_4+\gamma_5) + \gamma_1\gamma_3 J_{Cl}) + \gamma_5 J_{BClx+}(\gamma_1 J_{Cl} + \gamma_4 J_{BClx+})}$$

$$\theta_1 = \frac{\gamma_5 J_{BClx+}(\gamma_1 J_{Cl} + \gamma_4 J_{BClx+})}{\gamma_2\gamma_3 J_{BClx}^2 + J_{BClx}(\gamma_2 J_{BClx+}(\gamma_4+\gamma_5) + \gamma_1\gamma_3 J_{Cl}) + \gamma_5 J_{BClx+}(\gamma_1 J_{Cl} + \gamma_4 J_{BClx+})}$$

$$\theta_2 = \frac{J_{BClx}(\gamma_2 J_{BClx+}(\gamma_4+\gamma_5) + \gamma_1\gamma_3 J_{Cl})}{\gamma_2\gamma_3 J_{BClx}^2 + J_{BClx}(\gamma_2 J_{BClx+}(\gamma_4+\gamma_5) + \gamma_1\gamma_3 J_{Cl}) + \gamma_5 J_{BClx+}(\gamma_1 J_{Cl} + \gamma_4 J_{BClx+})}$$

The etching rate of GaN ($ER_{calc}$) and the polymer growth rate ($DR_{calc}$) are determined as follows:

$$ER_{calc} = \gamma_4 J_{BClx+}\theta_1 \frac{M(GaN)}{N_A\rho(GaN)}, \quad (7)$$

$$DR_{calc} = (\gamma_3 J_{BClx} - \gamma_5 J_{BClx+})\theta_2 \frac{M(B_xCl_y)}{N_A\rho(B_xCl_y)}, \quad (8)$$

where $M$ and $\rho$ are the molar mass and density of gallium nitride and the $B_xCl_y$ polymer, respectively, and $N_A$ is Avogadro's number. Figure 3 shows the experimentally measured GaN etching rate at values of $\langle\varepsilon_{ion}\rangle$ from Table 4, as well as the results of the analytical calculation of the etching rate ($ER_{calc}$) and the surface composition ($\theta$) with the corresponding fitting coefficients $\gamma_2 = \gamma_3 = 0.0002$, $\gamma_{04} = 0.5$, and $\gamma_{05}=1$. The absolute value of the polymer growth rate ($DR_{calc}$) was not calculated due to the unknown density, exact stoichiometric composition, and lack of experimental data on the growth rate. Therefore, in Figure 3, curve 2 qualitatively represents the polymer deposition process at ion energies $\langle\varepsilon_{ion}\rangle < \varepsilon_{trh}^{BxCly}$.

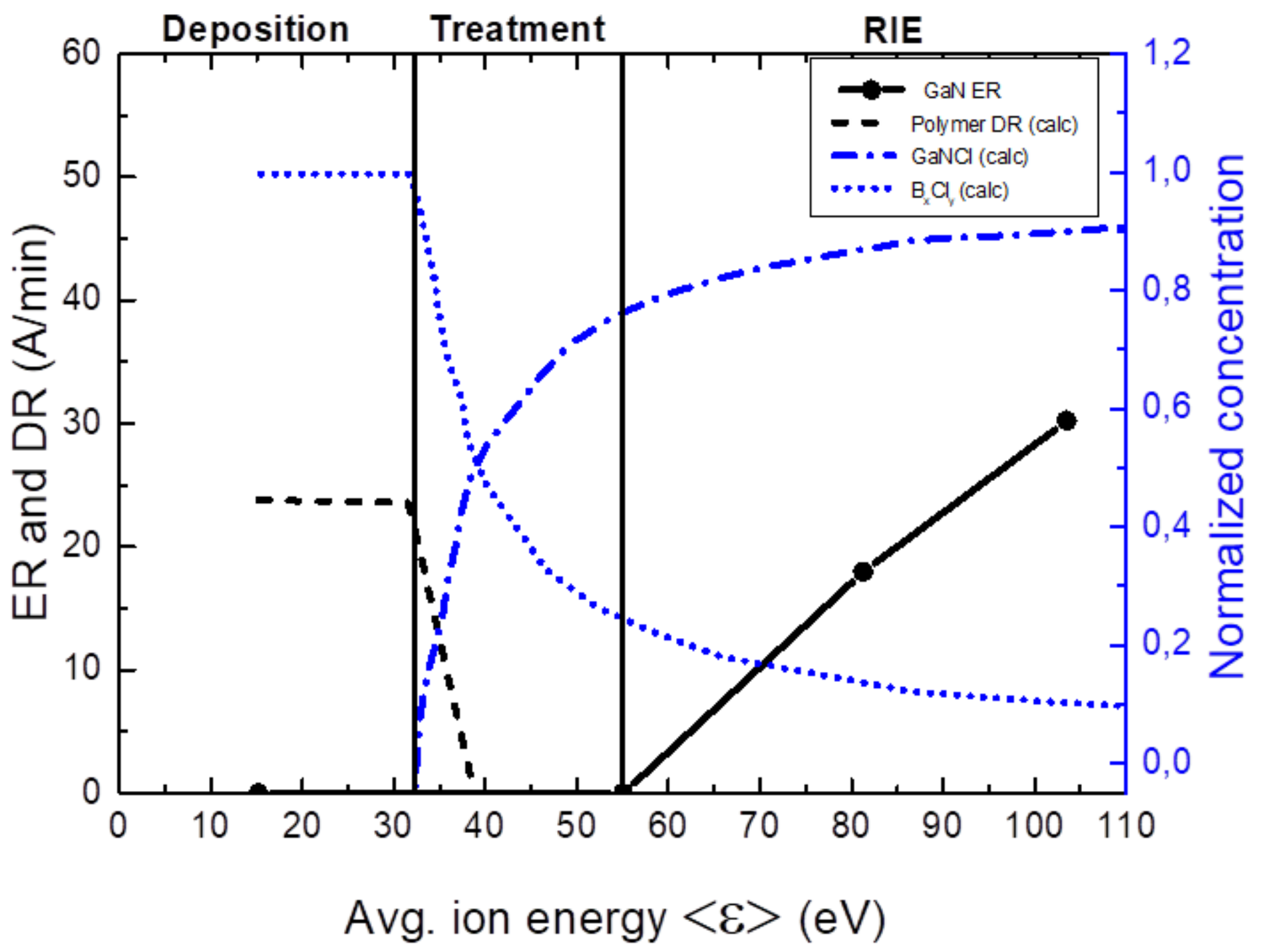

**Figure 3**. Dependence of the GaN etch rate (black solid line), $B_xCl_y$ growth (black dashed line) and surface composition (blue curves) on the average ion energy. Black circles – experimentally measured etch rate.

Figure 3 highlights three regimes of $BCl_3$ plasma interaction with GaN, depending on the energy of ions bombarding the surface ($\langle\varepsilon_{ion}\rangle$). Regime I ("Deposition") with $\langle\varepsilon_{ion}\rangle < \varepsilon_{trh}^{BxCly}$ corresponds to the regime of $B_xCl_y$ polymer film growth on the GaN surface. In this regime, the surface is completely covered by the polymer (Figure 3, blue doted curve). Regime II ("Treatment") in Figure 3 represents an intermediate regime of plasma surface treatment, where the normalized surface concentration of the polymer ($\theta_2$) decreases from 1 to 0.1 (Figure 3, blue doted curve). However, there is no etching of GaN (Figure 3, black solid line), as $\varepsilon_{trh}^{BxCly} < \langle\varepsilon_{ion}\rangle < \varepsilon_{trh}^{GaN}$. The regime corresponds to the experimental results at $U_{DC} = -20\ eV$, as shown in Table 3 and in reference [2]. Region III ("RIE"), with $\langle\varepsilon_{ion}\rangle > \varepsilon_{trh}^{GaN}$, represents the reactive ion etching regime of GaN. In this regime, the surface is activated by atomic chlorine (Figure 3, blue dash-dotted curve), and GaN is removed from the surface as GaCl and $N_2$ under ion bombardment. The fraction of $B_xCl_y$ compounds decreases with increasing $U_{DC}$, which qualitatively agrees with the measurement results [3].

## Conclusion

The calculations conducted in this work have shown that the ion energy from the $BCl_3$ plasma in the inductively coupled plasma (ICP) discharge is one of the key factors determining the transition between the reactive ion etching regime and the deposition regime of the $B_xCl_y$ polymer film on the surface of a solid body covered with an oxide layer. If, during the plasma treatment of GaN, the ion energies are below the threshold $\varepsilon_{trh}^{BxCly} \approx 32\ eV$, then strong chemical compounds such as $(BOCl)_3$, $B_3O_3Cl_2$, and $B_2OCl_3$ with oxygen cannot be removed from the surface under ion bombardment. Consequently, $BCl_x$ radicals from the plasma form chemical compounds like $B_xCl_y$ on the surface, leading to the growth of the polymer film.

Increasing the ion energy to values above the threshold $\varepsilon_{trh}^{GaN} \approx 60\ eV$ initiates the reactive ion etching of GaN, during which the material is removed from the surface as GaCl and $N_2$. Since the threshold $\varepsilon_{trh}^{GaN}$ and is higher than $\varepsilon_{trh}^{BxCly}$, an intermediate surface treatment regime for GaN exists. In this regime, the polymer and its compounds with oxygen are almost completely removed, and the surface is covered with atomic chlorine from the plasma.